\documentclass[aps,amsmath,amssymb,amsfonts,tightenlines,prd,10pt,onecolumn,preprint]{revtex4-1}

\usepackage{bm, slashed}
\usepackage{feynmp}
\usepackage[vcentermath,noautoscale]{youngtab}
\Yboxdim{8pt}

\newcommand{\fund}{\yng(1)}
\newcommand{\afund}{\overline{\yng(1)}}

\newcommand{\abfund}{\overline{\yng(1,1)}}

\newcommand{\Gc}{G_{\textrm{c}}}
\newcommand{\GF}{G_{\textrm{F}}}
\newcommand{\GFp}{G_{\textrm{F}}^\prime}
\newcommand{\UF}{\textrm{U}(1)_{\textrm{F}}}
\newcommand{\tlt}{\tilde\lambda_3}
\newcommand{\tlK}{\tilde\lambda_{K}}
\newcommand{\SU}{\textrm{SU}}
\newcommand{\UY}{\textrm{U}(1)_Y}
\newcommand{\Ua}{\textrm{U}(1)_a}
\newcommand{\UBL}{\textrm{U}(1)_{B-L}}

\begin{document}

\title{Composite Dirac Neutrinos}
\author{Yuval Grossman}
\author{Dean J. Robinson}
\affiliation{Laboratory for Elementary-Particle Physics, Cornell University, Ithaca, N.Y.}
\date{\today}

\begin{abstract}
We present a mechanism that naturally produces light Dirac
neutrinos. The basic idea is that the right-handed neutrinos are
composite. Any realistic composite model must involve `hidden flavor'
chiral symmetries. In general some of these symmetries may survive confinement,
and in particular, one of them manifests itself at low energy as an
exact $B-L$ symmetry. Dirac neutrinos are therefore produced. The
neutrinos are naturally light due to compositeness. In general, sterile
states are present in the model, some of them can naturally be warm
dark matter candidates.
\end{abstract}

\maketitle

\section{Introduction}
The possibility that various fields of the Standard
Model (SM) are composite has been considered in depth (see e.g. Refs
\cite{PDG:2008pd,Peccei:1983cm, Peskin:1981cq}). Apart from
potentially resolving the hierarchy problem, one of the key features
of composite theories is that they usually include a natural mechanism
to produce large fermion mass hierarchies. Put simply, the
bound-states, which form the degrees of freedom of the effective
low-energy theory, may acquire masses over a large range of scales via
a `tumbling' pattern of confinement-induced spontaneous chiral
symmetry breaking
\cite{DimopoulosRabySusskind:1980,RabyDimopoulosSusskind:1980tg,Napoly:1982ov}.

Compositeness can therefore explain the very small scale of the neutrino
masses. One particularly simple scenario was proposed in
Ref. \cite{ArkaniHamedGrossman:1999} and discussed in
Refs. \cite{GrossmanTsai:2008, Okui:2004xn}. The idea is to assume
that the right-handed neutrinos are massless chiral bound states produced
via the confinement of a hidden sector at a scale $\Lambda$. These
states acquire very small masses via interactions with the SM at a
much higher scale, $M \gg \Lambda$. The scale $M$ may be the
confinement scale of a yet more fundamental theory that condenses into
the SM and hidden sectors. After Higgs-induced electroweak spontaneous
symmetry breaking, the neutrino masses arise via irrelevant operators that are
suppressed by powers of $\Lambda/M$, and hence the neutrino masses are very light. In general one finds that both
light Dirac and Majorana mass terms can be produced for the neutrinos
in this manner.

A more popular mechanism used to produce light neutrino masses is the
see-saw mechanism, which requires the existence of heavy, sterile
Majorana neutrinos and produces light Majorana masses for the
neutrinos. As an alternative, it is interesting to consider whether
one can find a theory that naturally produces light \emph{Dirac}
neutrinos instead. Apart from compositeness, there are several known
mechanisms which can produce Dirac mass terms for the neutrinos. For
example, light Dirac masses can be produced via schemes involving
supersymmetry breaking \cite{Langacker:1998ut, DemirEtAl:2008dn,
Marshall:2009bk}, supergravity \cite{Abel:2004tt}, extra dimensions
\cite{Grossman:1999ra, ArkaniHamed:2001nm, Dienes:1998sb,
Hung:2002qp,Gherghetta:2003he}, discrete gauge symmetries
\cite{Davoudiasl:2005ks,Krauss:1989dg}, extra U(1) symmetries
\cite{Gogoladze:2001kj,Chen:2007gt}, or unparticles with large anomalous dimensions \cite{vonGersdorff:2008is}. In some of these cases Dirac
neutrinos naturally arise, whereas in others, Dirac
neutrinos are obtained by an \emph{ad hoc} imposition of lepton number.

In this paper, we explore the above mentioned compositeness scenario
further by considering the role of the symmetries of the hidden
sector, which naturally arise from the pattern of confinement-induced
spontaneous symmetry breaking of the preonic theory. Previous analyses
have not taken into account the fact that if one assumes that the
right-handed neutrinos are the chiral bound states in a confined hidden
sector, then the right-handed neutrinos must be non-trivially charged
under some chiral `hidden flavor' symmetry. These extra hidden flavor symmetries  
must feature in the structure of the neutrino mass terms. In particular, we show that there exists a
mechanism based on this hidden flavor symmetry that naturally produces
Dirac neutrinos. This mechanism arises in the following case: the
hidden flavor symmetry is a U(1) symmetry; the chiral bound states -- the
right-handed neutrinos -- all have the same hidden flavor charge; the
SM Higgs, $\phi$, is charged under this hidden flavor symmetry such that
the right-handed neutrinos may couple to the left-handed SM; and
$\phi$ is the only scalar in the low-energy theory with a non-trivial
vacuum expectation value (vev). The main result is that the U(1) axial
combination of the SM hypercharge and hidden flavor symmetries, which
is unbroken by $\langle \phi \rangle$, plays the role of lepton
number, guaranteeing that there are no Majorana masses. In other
words, instead of lepton number being either an accidental symmetry or
imposed \emph{ad hoc}, the hidden flavor symmetry -- necessarily
arising in our compositeness scenario from the chirality of the
right-handed neutrinos -- ensures that only Dirac neutrinos are
produced.

The paper is organized as follows. In Sec. \ref{sec:CCS}, we first
review the general mechanism through which composite right-handed
neutrinos may naturally produce light neutrino masses. In
Sec. \ref{sec:UHFM} we present a simple extension of the SM with
composite, chiral right-handed neutrinos and a U(1) hidden flavor
symmetry. Intriguingly, we show that once the right-handed SM fields are
also assigned an extra hidden flavor charge, then the theory is
non-anomalous.  Moreover, after spontaneous symmetry breaking, the
unbroken axial symmetry for this theory is isomorphic to $B-L$. The
theory therefore produces the usual SM Dirac fermions, along with
Dirac neutrinos that have heavily suppressed masses. Finally, in
Sec. \ref{sec:P} we examine the phenomenology of our low energy theory. The general structure of the neutrino mass
spectrum and the PMNS leptonic mixing matrix is presented. We also discuss
non-unitarity effects and possible dark matter candidates among the
heavier sterile neutrinos.

\section{Compositeness and Chiral Symmetry}
\label{sec:CCS}
\subsection{General dynamical framework}
\label{sec:DN}
We first provide a brief review of the underlying features and
assumptions inherent to composite neutrino models. More details can be
found, for example, in Ref.~\cite{tHooft:1980, Weinberg:1996,
Peccei:1983cm, Peskin:1981cq}.

In the study of composite theories, one generally posits the existence
of an ultraviolet theory of chiral fermions, called preons, which has
a chiral gauge $\otimes$ flavor symmetry. This theory undergoes
confinement to produce a low-energy effective theory whose degrees of
freedom are bosonic and fermionic bound states. These bound states are singlets of the original gauge group, but may have
non-trivial flavor symmetries.

The formation of such singlets is just a mathematical exercise that is
subject to the 't~Hooft anomaly matching conditions \cite{tHooft:1980, Weinberg:1996}. The main difficulty in
composite theories is to determine the dynamics of confinement, which
in turn determines the physical bound states, the symmetries of
the effective theory, and the scales at which confinement occurs.  For
example, confinement could occur at just one scale. Alternatively one
could conceive of a preonic theory which first condenses into an
effective theory with weak, asymptotically-free
effective couplings, so that at some lower scale it further condenses
into a yet lower energy effective theory. At each scale, scalar condensates
could induce spontaneous symmetry breaking, so that the final unbroken
symmetries and physical degrees of freedom for these two cases may be
different.

One approach determines this dynamical information via
the complementarity principle. In brief, the complementarity principle
is the idea that the pattern of confinement-induced symmetry breaking
and the chiral bound states can be determined from a dynamical
symmetry breaking scheme for the preonic theory in the absence of
confinement. An exposition of this principle and other necessary
assumptions can be found elsewhere (see e.g. Refs
\cite{DimopoulosRabySusskind:1980,RabyDimopoulosSusskind:1980tg,Napoly:1982ov}). For
our purposes it is sufficient to note the following:
\begin{enumerate}
	\item Confinement may occur in stages at successively lower energy scales. At each scale, bound states with non-trivial gauge and flavor charges may be produced, and there may exist a scalar condensate with a non-trivial vacuum expectation value, which leads to spontaneous symmetry breaking of the chiral symmetry. The original gauge $\otimes$ flavor symmetries are thus said to `tumble' down through these stages to a final unbroken subgroup. Tumbling and confinement ceases when one ends up with a low-energy effective field theory for which the chiral bound states are all confining gauge singlets.
	\item The original chiral symmetry may or may not be completely spontaneously broken by confinement. At any stage of confinement, if the chiral symmetry is not completely broken, then the 't~Hooft anomaly matching constraints imply that there must be chiral, massless bound states in the effective theory produced below the corresponding confinement scale.
	\item At a confinement scale $\Lambda$, fermionic bound states furnishing real representations may acquire either Majorana or Dirac masses, with masses typically at the confinement scale.
	\item However, if these fermionic bound states are gauge singlets and interact with scalar condensates only via the heavy gauge bosons or scalars produced at some higher scale $\Lambda^\prime \gg \Lambda$, then their mass will be generated at loop level, and will be suppressed by at least a factor $(\Lambda/\Lambda^\prime)^2$. This is the so-called secondary mass generation mechanism \cite{DimopoulosRabySusskind:1980,Peccei:1983cm}, which is the similar to the mechanism by which the fermions acquire masses in extended technicolor theories.
\end{enumerate}

\subsection{Generation of light masses}
\label{sec:GLM}
This section recapitulates the ideas of Ref. \cite{ArkaniHamedGrossman:1999}. The general approach of composite
neutrino models is to suppose that the right-handed neutrinos are the chiral bound states of a hidden sector, which condenses at scale $\Lambda$
but couples to the SM via some higher scale $M \gg \Lambda$. The idea is that the compositeness of the right-handed neutrinos suppresses their effective
Yukawa coupling to the SM by powers of $\Lambda/M$.

Let us suppose that there exists a preonic theory which undergoes confinement at a scale $M$, such that the gauge $\otimes$ flavor symmetry groups of the theory spontaneously break down to a subgroup $\Gc\otimes\GF\otimes G_{\textrm{SM}}$, where $\Gc$ ($\GF$) is a confining gauge (flavor) symmetry and $G_{\textrm{SM}}$ is the usual $\SU(3)_c\otimes \SU(2)_L\otimes \UY$ symmetry of the Standard Model. For the sake of clarity later on, we will henceforth call the degrees of freedom of the original preonic theory \emph{UV preons}.

Let us suppose that below the confinement scale $M$ we have an effective theory with: (1) the usual left-handed leptonic SM fields, $L_L$, which are $\Gc\otimes\GF$ singlets; (2) chiral bound states $q$, which are SM singlets but furnish non-trivial $\Gc\otimes\GF$ representations; (3) a scalar condensate $\phi$ which is a $\Gc$ singlet, but has non-trivial $\GF$ charges and otherwise has the charges of the SM Higgs. The chiral bound states $q$ act as preons for the $\Gc\otimes\GF$ theory, so henceforth we refer to them as \emph{effective preons}. As they are SM singlets, we say that the effective preons $q$ comprise a hidden sector. It follows from this hypothesis that the SM and hidden sectors may only interact via the exchange of heavy messengers at scale $M$.

Now let us suppose that there exists a combination of $n$ effective preons, crudely denoted by $q^n$, that contains a right-handed spin-1/2
Lorentz representation, and has precisely the correct flavor charges such that $\phi^* q^n$ is a $\GF$ singlet. Note that since $q^n$ is spin-1/2, then $n$ is odd and $n\ge3$. 
Integrating out heavy degrees of freedom, we have an effective irrelevant vertex
\begin{equation}
	\label{eqn:LYSM}
	\mathcal{L}_{\textrm{yuk}} = \frac{\lambda}{{M}^{3(n-1)/2}}\bar{L}_L\tilde{\phi}q^n~,
\end{equation}
where $\lambda$ is an $\mathcal{O}(1)$ number and as usual $\tilde\phi_a = \epsilon_{ab}\phi^{b*}$ with respect to $\SU(2)_L$ indices. Well below confinement scale $M$, the hidden sector suffers further confinement, possibly over multiple scales 
\begin{equation}
	M \gg \ldots \gg \Lambda_i \gg \ldots \gg \Lambda~,
\end{equation}
such that confinement ceases below the $\Lambda$ scale. For the sake of simplicity, hereafter we assume that the hidden sector suffers confinement as just one scale $\Lambda$. Therefore,
at scale $\Lambda$, $q^n$ condenses into a right-handed
spin-1/2 bound state $N_R$
\begin{equation}
	\label{eqn:HSPC}
	 q^n \to N_R \Lambda^{3(n-1)/2},
\end{equation}
and we end up with low-energy effective Yukawa
\begin{equation}
	\label{eqn:GCSEL}
	\mathcal{L}_{\textrm{yuk}} = \lambda\bigg(\frac{\Lambda}{M}\bigg)^{\frac{3(n-1)}{2}} \bar{L}_L\tilde{\phi}N_R~,
\end{equation}
where $\lambda$ is some redefined $\mathcal{O}(1)$ coupling. As advertised, the compositeness of $N_R$ has suppressed the Yukawa coupling in Eq. (\ref{eqn:GCSEL}) by powers of 
\begin{equation}
	\epsilon \equiv \frac{\Lambda}{M} \ll 1~.
\end{equation}
After $\phi$ acquires a non-trivial vev, then neutrino masses are produced
from Eq.~(\ref{eqn:GCSEL}). Since $n \ge 3$, such masses
will be at most of the order $\langle \phi \rangle \epsilon^{3}$. For
sufficiently small $\epsilon$, the neutrino Dirac masses will be
suppressed compared to those of the charged leptons. Note also that
the larger the number of effective preons in $N_R$, the greater the suppression.

The Yukawa term in Eq. (\ref{eqn:GCSEL}) is not the only possible mass generating term. As yet there is no symmetry which prevents the formation of Majorana mass terms from higher dimensional operators. As an example, the operator
\begin{equation}
	\label{eqn:HDOH}
	\frac{1}{M}(L_L\tilde{\phi})^T\sigma^2L_L\tilde{\phi}~,
\end{equation}
produces a Majorana mass for the left-handed neutrinos.  In summary so far, in this class of composite neutrino models, one finds both light Dirac and Majorana mass terms.

\subsection{Renormalization Effects}
In our discussion so far we have omitted possible effects due to renormalization group (RG) running. For example, Eq. (\ref{eqn:GCSEL}) involves vevs evaluated at the two different scales: The first is due to the condensation of the UV preons that breaks the UV preonic theory down to $\Gc\otimes\GF\otimes G_{\textrm{SM}}$ at scale $M$; the second arises from the condensation of the effective preons in the hidden sector at scale $\Lambda$. To be consistent in choice of renormalization scale, it is therefore necessary to run the operator (\ref{eqn:LYSM}) from scale $M$ down to scale $\Lambda$, so we should expect that RG effects will modify Eq. (\ref{eqn:GCSEL}). However, the tumbling scenario outlined in Sec. \ref{sec:DN} implictly assumes that the effective theory between scales $M$ and $\Lambda$, of which (\ref{eqn:LYSM}) is an operator, is asymptotically free. As a result, we expect RG effects to introduce power-logarithmic corrections to Eq. (\ref{eqn:GCSEL}).

An alternative is to consider the possibility that the effective theory of effective preons $q$ is approximately conformal and strongly coupled between scales $M$ and $\Lambda$. In this case we can contemplate large, constant anomalous dimensions for operators such as (\ref{eqn:LYSM}) or for the loop-generated masses produced by the secondary mass generation mechanism. This is similar to the well-known walking technicolor scenario (see e.g. Ref. \cite{Hill:2002ap} for a review). However, this strong, conformal scenario is inconsistent with the tumbling scenario, because the confinement scale of the effective preons becomes badly defined if they have strong, approximately conformal dynamics above $\Lambda$. (We note that the idea of conformal right handed neutrinos that couple to the SM through irrelevant operators has been recently considered in Ref. \cite{vonGersdorff:2008is}.) 

We expect the power-logarithmic RG effects in the asymptotically free, tumbling scenario will lead to $\mathcal{O}(1)$ or smaller corrections to Eq. (\ref{eqn:GCSEL}), depending on the details of the effective preonic theory in the hidden sector. In this paper, we seek to describe only the model-independent effects of the hidden sector on mass and coupling scales (up to assumptions about its symmetry breaking pattern), so these corrections can be absorbed into the parameter $\lambda$, which we assumed was just an $\mathcal{O}(1)$ number. In other words, while RG effects could lead to important corrections in a specific model, they do not change the scale of the Yukawa coupling to the right-handed neutrinos, and we therefore need not consider RG effects henceforth in this paper.

\subsection{Role of chiral symmetry}
The confinement down to the scale $\Lambda$ generally results in the spontaneous symmetry breaking of $\Gc\otimes\GF \to \Gc^\prime\otimes\GFp$. Since we assumed that confinement stops below $\Lambda$, it must be that any chiral bound states are $\Gc^\prime$ singlets, but furnish complex $\GFp$ representations. 

Crucial to the above analysis leading to Eq. (\ref{eqn:GCSEL}) is the
implicit assumption that the $N_R$ are a subset of these chiral
bound states. If this was not the case, then they would acquire Dirac or
Majorana mass terms at scale $\Lambda$ or higher. Hence, if the $N_R$ are to be chiral, then some chiral flavor symmetry $\GFp$ must survive confinement. Henceforth
we call $\GFp$ the hidden flavor symmetry. Note that the 't~Hooft anomaly matching conditions require that the $\GFp$ anomalies of the chiral bound states match those of the original UV preons.

This chiral hidden flavor symmetry has two important consequences. First, as in the 't~Hooft formalism, the chiral symmetry ensures that there must be elementary spectator fermions, whose $\GFp$ anomalies cancel those of the chiral bound states.  The second consequence is that the scalar $\phi$ must also have hidden flavor charges: since $\phi^*N_R$ is a $\GFp$ singlet, and since $N_R$ transforms under a complex representation of $\GFp$, then it follows that $\phi$ transforms non-trivially under $\GFp$ too. 

\subsection{Dirac neutrinos}
\label{sec:DIN}

Let us now present a toy model that produces light Dirac neutrinos. The main idea is that inclusion of
the chiral hidden flavor symmetry together with some special choices will result in an unbroken lepton number.

Suppose $\GFp$ is just a U(1) symmetry
\begin{equation}
	\GFp \equiv \UF~,
\end{equation}
so that $N_R$ and $\phi$ both have the same non-zero $\UF$ charge, which we denote hereafter by $\gamma$. We then have $\SU(2)_L\otimes \UY\otimes \UF$ group structure as shown in Table \ref{tab:GSMM}.
\begin{table}[t]
\begin{tabular}{c|cccc}
	\hline
	Field & $\quad \SU(2)_L\quad$ & $\quad \UY\quad$ &  $\quad \UF\quad$ & $\quad \Ua\quad$\\[3pt]
	\hline\hline
	$\phi$ & $\fund$ & $\frac{1}{2}$ & $\gamma$& 0\\[3pt]
	\hline
	$L_L$ & $\fund$ & $-\frac{1}{2}$ & 0 & $-\gamma/2$\\[3pt]
	$N_R$ & $\bm{1}$ & 0 & $\gamma$ & $-\gamma/2$\\[3pt]
	\hline
\end{tabular}
\caption{SM, hidden flavor, and axial representations of the $\UF$ model of
section~\ref{sec:DIN}.}
\label{tab:GSMM}
\end{table}
Now, observe that $\phi$ is uncharged under the U(1) axial combination
of the $\UY$ and $\UF$, which has charge $a \equiv \gamma Y - F/2$ and
is denoted by $\Ua$. (Here and henceforth $Y$ ($F$) denotes the $\UY$
($\UF$) charge of the field in question.) The axial symmetry $\Ua$ is
therefore unbroken by $\langle \phi \rangle$, and so it remains a
symmetry of the spontaneously broken theory. Moreover both $L_L$ and
$N_R$ have axial charge $a =-\gamma/2$. Hence for the field content of
Table \ref{tab:GSMM}, global $\Ua$ is isomorphic to lepton number. It
immediately follows if $\phi$ is the only scalar which gets a
vev, then Majorana masses cannot be produced by spontaneous symmetry
breaking: only Dirac masses are produced, and these are light due to
compositeness, as in Eq. (\ref{eqn:GCSEL}). Note that there may be
other higher dimensional operators apart from the Yukawa term in
Eq. (\ref{eqn:GCSEL}) that also produce contributions to the Dirac
masses. However, these contributions will be further suppressed by
powers of $\langle \phi \rangle/M$ and $\epsilon$.

\section{A U(1) hidden flavor model}
\label{sec:UHFM}
\subsection{Field Content}
\label{sec:UOFC}
We now seek to exploit the above result to produce a realistic
extension of the SM with light Dirac neutrinos. Following from the
above, we choose $\GFp \equiv \UF$. This choice has the added
advantage that $\phi$ has the same number of field degrees of freedom
as the SM Higgs. Let us now suppose the effective preonic
$\Gc\otimes\GF$ theory can be chosen such that there are exactly three
chiral right-handed bound states, $N_R^i$, all with the same $\UF$ charge,
$\gamma$.  Examples of such theories are presented in Appendix
\ref{sec:EPT}.

We assume that $\phi$ has the same charges as in Table \ref{tab:GSMM} and is the only scalar which acquires a vev. It follows from this assumption that in order for the usual SM mass structure to be produced, we must have Yukawa terms of the form
\begin{equation}
	\mathcal{L}_{\textrm{yuk}} = \bar{L}_L^i\phi E_R^j + \bar{L}_L^i\tilde{\phi}N_R^j + \bar{Q}_L^i\phi D_R^j + \bar{Q}_L^i\tilde{\phi}U_R^j + \mbox{h.c.}~,
\end{equation}
where $E_R$, $D_R$, $U_R$ and $Q_L$ are the right-handed charged leptons, down and up quarks and left-handed quark doublets respectively. (We henceforth refer to $E_R$, $U_R$ and $D_R$ as the right-handed SM fields.) Assuming that the left-handed SM fields have no hidden flavor charges, then in order for such terms to exist the right-handed SM fields must also have $\UF$ charges. The only possible hidden flavor charge assignments are shown in Table \ref{tab:HFMC2}, along with the usual SM charges.

\begin{table}[t]
\begin{tabular}{c|ccccc}
	\hline
	Field & $\quad \SU(3)_c$ & $\quad \SU(2)_L$ & $\quad \UY$ &  $\quad \UF$ & $\quad \Ua$ \\[3pt]
	\hline\hline
	$\phi$  &$\bm{1}$& $\fund$ & $\frac{1}{2}$ & $\gamma$ & 0\\[3pt]
	\hline
	$L_L^i$ &  $\bm{1}$ & $\fund$ & $-\frac{1}{2}$ & 0 & $-\frac{\gamma}{2}$\\[3pt]
	$E_R^{*i}$ &$\bm{1}$& $\bm{1}$ & 1 & $\gamma$ & $\frac{\gamma}{2}$\\[3pt]
	$N_R^{*I}$ &$\bm{1}$& $\bm{1}$ & 0 & $-\gamma$ & $\frac{\gamma}{2}$\\[3pt]
	$N_L^{\alpha}$ &$\bm{1}$& $\bm{1}$ & 0 & $\gamma$ & $-\frac{\gamma}{2}$\\[3pt]
	$Q_L^i$ & $\fund$ & $\fund$ & $\frac{1}{6}$ & 0 & $\frac{\gamma}{6}$\\[3pt]
	$U_R^{*i}$ & $\afund$ & $\bm{1}$ & $-\frac{2}{3}$ & $-\gamma$ & $-\frac{\gamma}{6}$\\[3pt]
	$D_R^{*i}$ & $\afund$ & $\bm{1}$ & $\frac{1}{3}$ & $\gamma$ & $-\frac{\gamma}{6}$\\[3pt]
	\hline
\end{tabular}
\caption{Scalar and left-handed fermionic field content for the $U(1)$
hidden flavor model describes in section~\ref{sec:UOFC}. As above, the axial charge $a = \gamma Y - F/2$. It is clear that $\Ua \simeq \UBL$.}
\label{tab:HFMC2}
\end{table}

It is intriguing that the unbroken axial symmetry in this theory is isomorphic to $B-L$. Hence $B-L$ remains an unbroken symmetry in this theory, and only Dirac fermions are produced by the spontaneous symmetry breaking induced by $\langle \phi \rangle$. This is one of the main results of this paper. 

Before proceeding, several comments are in order. First, there may be an arbitrary number of right-handed bound states with the hidden flavor charge $\gamma$: we have denoted these as $N_R^I$. In line with the above supposition, a subset of three of these bound states, denoted by $N_R^i$, are chiral. The remainder, denoted $N_R^\alpha$, must form massive Dirac fermions with left-handed bound states $N_L^\alpha$, which must therefore have $\UF$ charge $\gamma$. These Dirac neutrinos will typically have masses at the $\Lambda$ confinement scale. (It is possible, however, that some of these masses are suppressed via the secondary mass generation mechanism.) We have implicitly adopted here the following index notation, which we will continue to use throughout the remainder of this paper: Upper case Roman indices denote all spin-1/2 bound states of a particular $\UF$ charge; lower case Roman indices denote chiral bound states and leptonic flavor; lower case Greek indices denote massive spin-1/2 bound states. 

Second, the $\UF$ charges in Table \ref{tab:HFMC2} are clearly commensurate, which permits this U(1) to be embedded in a semi-simple Lie group.  This must be the case, since $\UF$ was generated by spontaneous symmetry breaking of a larger group.

Finally, note that in principle there may be various other SM sterile spin-1/2 bound states in the theory with hidden flavor charge $F \not= \pm\gamma$. By hypothesis none of these are chiral, so they must be heavy with masses generally at scale $\Lambda$. More significantly, these bound states necessarily have axial charge $a \not=\mp\gamma/2$, so that they cannot form Dirac mass terms with the neutrinos. We therefore neglect them henceforth.

\subsection{Anomaly cancellation}
A crucial issue is the cancellation of the anomalies. It is clear that
there are no anomalies in the SM sector, but there may be non-trivial
anomalies involving $\UF$. Let us therefore examine all these
anomalies. Clearly there is no $\SU(2)_L^2\UF$ anomaly since the
left-handed SM is not charged under $\UF$. Let $N$ be the number of
physical right-handed neutrinos, i.e. let $I = 1,\ldots, N$, so that
$\alpha = 4,\ldots,N$. Then we have anomalies
\begin{align}
	&\mathcal{A}\big[\UF^3\big] 
	 \propto \big[3 - N + (N-3) + 3 - 3\big]\gamma^3  =0~,\notag\\
	&\mathcal{A}\big[\mbox{gravity}^2\UF\big] 
	\propto \big[3 - N + (N-3) + 3 - 3\big]\gamma =0~,\notag\\
	&\mathcal{A}\big[\SU(3)_c^2\UF\big]
	 \propto (3\gamma - 3\gamma)  = 0~,\notag\\
	&\mathcal{A}\big[\UY^2\UF\big] 
	 \propto \gamma - 3\gamma\bigg(-\frac{2}{3}\bigg)^2 + 3\gamma\bigg(\frac{1}{3}\bigg)^2   = 0~,\notag\\
	&\mathcal{A}\big[\UY\UF^2\big]
	 \propto \gamma^2 -3\frac{2}{3}\gamma^2 + 3\frac{1}{3}\gamma^2   = 0~.
\end{align}
So with the above hidden flavor assignments all the anomalies involving $\UF$ cancel. We have thus shown that the theory presented in Table \ref{tab:HFMC2} is a consistent extension of the SM with composite Dirac neutrinos.  

\subsection{Right-handed SM}
So far we have not discussed the compositeness of the right-handed
SM. In our model the right-handed SM is charged under $\UF \subseteq
\GF$, so we would na\"\i vely expect these fields to be composite at
scales $\Lambda$ (or $\Lambda_i$). The SM Yukawa couplings would then be
suppressed by  $\epsilon$ (or $\Lambda_i/M$, which might account
for inter-family mass splittings, but we do not consider this possibility
further in this paper). 

It is also conceivable that the field content and group structure of
the UV preonic theory can be chosen such that these fields are composite
only at scale $M$. In this scenario the right-handed SM Yukawa
couplings are not suppressed by compositeness. The reason is that
these Yukawa terms now depend on only one compositeness scale, $M$,
which must cancel by na\"\i ve dimensional analysis. In comparison,
the compositeness of the neutrinos at scale $\Lambda$ suppresses their
Yukawa couplings to the left-handed leptons by powers of $\epsilon$,
as in Eq. (\ref{eqn:GCSEL}). As a result, the large mass hierarchy
between the neutrinos and the rest of the SM fermions is achieved.

With reference to the 't~Hooft anomaly matching formalism, another possibility is to identify the right-handed SM as spectators, that is, as the elementary fields which are uncharged under any of the confining gauge groups, but which precisely cancel the `flavor' $\SU(3)_c\otimes \UY\otimes\UF$ anomalies.  In this case, the SM Yukawa couplings are similarly not suppressed by compositeness. A further advantage of this scenario is that the compositeness contributions to e.g. the anomalous magnetic moment are suppressed, such that the strong bounds on $M$ due to the electron $g-2$ are evaded.

\subsection{Symmetry breaking pattern and gauge bosons}
Let us now proceed to further examine the pattern of symmetry breaking for this theory.  Since gauge symmetries are not violated by quantum gravity effects, we assume $\UF$ is a weakly gauged symmetry. This assumption also follows from the usual 't~Hooft prescription.

The effective Higgs $\phi$ has the same number of degrees of freedom
and same scalar potential $V(\phi)$ as in the SM. We are therefore
free to choose the usual unitarity gauge defined by $\langle \phi
\rangle = (0,v)^T$. In this gauge it is clear that the electromagnetic
symmetry $U(1)_{\textrm{EM}}$, with generator $Q = T^3 + Y$, remains a
gauge symmetry of the theory. Further, it is clear that $\Ua \simeq
\UBL$ is a gauge symmetry too.

Apart from the $\SU(3)_c$ generators, there are two unbroken generators after spontaneous symmetry breaking. These generators must correspond to two U(1)s, whose generators must be linearly independent combinations of $Q$ and $B-L$. We may therefore write the electroweak symmetry breaking pattern for this theory as
\begin{equation}
	\SU(2)_L \otimes \UY \otimes \UF \to U(1)_{\textrm{EM}}\otimes U(1)^\prime~,
\end{equation}
where the generator of $U(1)^\prime$ is a linear combination of $Q$ and $B-L$, and its gauge boson, $A^\prime_\mu$, is orthonormal to the photon $A_\mu$.  An explicit presentation of the gauge boson mass basis and couplings is presented in Appendix \ref{sec:GBSC}. 

Let us now suppose that $\UF$ is weakly gauged compared to the
SM. That is, we define
\begin{equation}
	\kappa  \equiv \frac{2\gamma g_{\textrm{F}}}{\sqrt{g^2 + g^{\prime 2} + (2\gamma g_{\textrm{F}})^2}}
\end{equation}
where $g_{\textrm{F}}$ is the gauge coupling of the $\UF$, and we
assume $\kappa \ll 1$. This is a reasonable assumption if $\UF$ is a
subgroup of the UV preonic flavor group, while $G_{\textrm{SM}}$ has
generators arising from the UV preonic confining gauge group. For our
present purposes, it is sufficient to note that in the $g_F \ll g$
limit we recover the usual SM gauge boson structure along with a
$A_\mu^\prime$ gauge boson that is weakly coupled to all fields. In
particular, the covariant derivative in this limit (\ref{eqn:ACDWGL})
is
\begin{align}
	iD_\mu
	& \simeq i\partial_\mu -gT^{\pm}W^{\pm}_\mu  - eQA_\mu - \frac{g}{c_W}\bigg[ \big(T^3 - Qs_W^2\big) - \frac{\kappa^2}{2}\big(Qc_W^2 +Y - B-L\big)\bigg]Z_\mu\notag\\
	& - \frac{g\kappa}{c_W}\bigg[Qc_W^2 - \frac{B-L}{2}\bigg]A^\prime_\mu~,\label{eqn:CDWGL}
\end{align}
for which all terms are defined in Appendix
\ref{sec:GBSC}.  Note that the couplings of the $A_\mu^\prime$ to both
the SM and hidden sector is suppressed by a factor $\kappa$, while the
coupling of the hidden sector to the $Z$ is even more strongly
suppressed by a factor $\kappa^2$. In general, the phenomenological
consequences of this extra gauge boson are suppressed by $\kappa$,
which we can always choose small enough to satisfy experimental
bounds.

In principle, kinetic mixing between the photon and $A_\mu^\prime$ can occur in this theory. However, any such mixing is suppressed by $\kappa$. Moreover,  by construction the gauge group of this theory embeds into the gauge $\otimes$ flavor groups of a fundamental UV preonic theory. If one assumes the UV preonic theory contains only a single U(1) factor, then the underlying symmetry of the UV preonic theory prevents kinetic mixing between orthonormal U(1) gauge bosons in the full quantum low energy theory.

\section{Phenomenology}
\label{sec:P}
We now proceed to examine some phenomenological aspects of our U(1)
hidden flavor model.  The
parameters characterizing compositeness are just $\Lambda$ and
$\epsilon$. We emphasis that, by construction, this model is an
effective low-energy theory, which is well-defined only at scales well
below the $\Lambda$ confinement scale. Hence the phenomenology
presented in the following section holds only at scales much less than
$\Lambda$.

Cosmological and in particular big-bang nucleosythesis constraints on $\Lambda$ are presented in Ref. \cite{Okui:2004xn}.  In the case that the hidden sector undergoes chiral symmetry breaking, as we have assumed throughout, one requires $\Lambda \gtrsim r^{1/3}$ GeV, where $r$ is the ratio of right and left handed neutrino number densities. It is therefore expected, though not necessary, that $\Lambda$ should be greater than the GeV scale.  

In general, the phenomenology of this model is the same as that of the
well-known SM with Dirac neutrinos, up to corrections due to
compositeness and the $A_\mu^\prime$ coupling. The former results in
non-diagonal couplings of the neutrinos to the $Z$ and Higgs as well
as a non-unitary PMNS matrix. The latter amounts to a rescaling of the
couplings. However, all these corrections prove to be strongly
suppressed by $v/M$, $\epsilon$ and/or $\kappa$ factors. That is, so long as these factors are
sufficiently small, the phenomenology of exotic processes is
indistinguishable from the SM with Dirac neutrinos within current
experimental precision. The spirit of this section is to verify this suppression explicitly.

\subsection{Baryon and lepton number violation}
Since only $B-L$ is an exact symmetry of our theory, higher dimensional $B$ and $L$ violating operators are in principle permitted in the low energy theory. Hence proton decay and deuterium decay can occur, although neutrinoless double-$\beta$ decay is forbidden. The exact structure of any particular $B$ violating operator depends on the details of the UV preonic theory. However, we can estimate upper bounds on any particular $B$ or $L$ violating process from the low energy effective field theory, by assuming any such operator exists at the lowest possible mass dimension. Note that since we assumed that the SM was composite at scale $M$, then any $B$ violating process must be mediated by heavy gauge bosons at scale $M$ or higher. 

For example, the simplest operator which produces the proton decay $p \to e^+\gamma$ has the form $uude/M^2$ by na\"\i ve dimensional analysis. Hence in our theory the proton decay rate is suppressed by at least $(\Lambda_{\textrm{QCD}}/M)^4$. Similarly, we expect any $B$ violating process to be suppressed by powers of $\Lambda_{\textrm{QCD}}/M$. The current proton decay bound \cite{PDG:2008pd} implies that $M \gtrsim 10^{15}$ GeV.  In contrast, $M \gtrsim 10^3$ TeV is well-motivated by suppression of flavor changing neutral currents in extended technicolor models (see e.g. Ref. \cite{Hill:2002ap}), while bounds from four-Fermi contact interactions typically require $M \gtrsim 10$ TeV \cite{PDG:2008pd}. Generally, these discrepancies are resolved by detailed consideration of the structure of the UV preonic theory, which is beyond the scope of the present paper. (A detailed generic discussion regarding bounds on the SM compositeness scale can be found in e.g. Ref. \cite{Peccei:1983cm}.) For the purposes of this paper, we will henceforth assume just that $M \gtrsim 10$ TeV, and that tighter experimental bounds are satisfied by the details of the UV preonic theory.

\subsection{Yukawa terms}
The leading order (in $v/M$) mass generating terms for the leptonic sector of the U(1) hidden flavor theory are simply the Yukawa terms
\begin{align}
	\lambda_{ij}^{\ell}\bar{L}_L^i\phi E_R^j + \lambda_{iI}\epsilon^{3(n_I-1)/2}\bar{L}_L^i\tilde{\phi} N_R^I + \bar{N}_L^\alpha \Lambda_\alpha N_R^\alpha~, \label{eqn:LHFM}
\end{align}
where $n_I$ is the number of effective preons comprising $N_R^I$, $\lambda$ and $\lambda^{\ell}$ are $\mathcal{O}(1)$ matrices, and we have written the massive spin-1/2 bound states in their mass basis, with masses $\Lambda_\alpha \sim \Lambda$ (or $\sim\Lambda \epsilon^{2}$ if there is any secondary mass generation). 

Let us now suppose that there are in total $N$ spin-1/2 bound states which correspond to physical right-handed neutrinos, of which three are chiral and $K \equiv N-3$ are massive. Further, let us also redefine
\begin{equation}
	\epsilon^{3(n_I -1)/2}\lambda_{iI} \equiv \epsilon^{\tilde n}\tilde\lambda_{iI}
\end{equation}
where $\tilde{n} = \mbox{min}_I[3(n_I -1)/2]$ and $\tilde\lambda$
absorbs the remaining powers $\epsilon$.  That is, we factored out the
largest possible power of $\epsilon$ to form a $3\times N$ matrix,
$\tilde\lambda$, with at least one $\mathcal{O}(1)$ column: The
remaining entries are suppressed by powers of $\epsilon$. After
spontaneous symmetry breaking we write the mass terms as
\begin{equation}
	\label{eqn:LLY}
	\mathcal{L}_{\textrm{m}} = v\lambda_{ij}^{\ell}\bar{E}_L^i E_R^j + \Lambda \begin{pmatrix} \bar{N}_L^i & \bar{N}_L^\alpha \end{pmatrix} A \begin{pmatrix} N_R^i \\ N_R^\alpha \end{pmatrix} + \mbox{h.c.},
\end{equation}
where $A$ is an $N\times N$ square matrix
\begin{equation}
	\label{eqn:NMM}
	A  \equiv \begin{pmatrix} \theta\tlt & \theta\tlK \\ 0 & d_K\end{pmatrix}~.
\end{equation}
Here the parameter
\begin{equation}
	\label{eqn:DMALH}
	\theta \equiv \frac{v}{\Lambda}\epsilon^{\tilde n}~, 
\end{equation}
the subscripts denote the number of columns, and $d_K$ is diagonal with entries $\Lambda_\alpha/\Lambda$. We assume that mostly $[d_K]_{\alpha\alpha} \sim \mathcal{O}(1)$, but there may exist some entries of $d_K$ which are suppressed by $\epsilon^2$ due to secondary mass generation.  

\subsection{Leptonic mixing matrix}

In general, both $\lambda^{\ell}$ and $A$ have biunitary decompositions of the form 
\begin{equation}
	\label{eqn:BUD}
	\lambda^{\ell}  = V^{\ell}d^{\ell}W^{\ell\dagger}~,~~\mbox{and}~ A  = V d W^{\dagger}~, 
\end{equation}	
where $d^{\ell}$ ($d$) is a $3\times 3$ ($N\times N$) diagonal matrix and $V^{\ell}$ and $W^{\ell}$ ($V$ and $W$) are $3\times3$ ($N\times N$) unitary matrices. As usual we define the lepton mass basis by
\begin{align}
	e_L & = V^{\ell\dagger}E_L & e_R&=W^{\ell\dagger}E_R\notag\\
	\nu_L & = V^{\dagger}N_L & \nu_R &= W^{\dagger}N_R~.\label{eqn:MBN}
\end{align}
A notable difference from the SM is that here there are $3$ charged lepton mass eigenstates, but $N$ neutrino mass eigenstates. 

In the neutrino mass basis the $W$ couplings are non-diagonal, as usual. Since the $N_L^i$ and $N_L^\alpha$ have different $Z$ couplings, as can be seen in Eq. (\ref{eqn:CDWGL}), $Z$ neutrino couplings in the mass basis are non-diagonal too. To order $\mathcal{O}(\kappa^2)$, one finds that non-diagonal neutrino currents are
\begin{align}
	\mathcal{J}_W^{\mu-} &=  \frac{g}{\sqrt{2}}\bar{e}_LU\gamma^\mu \nu_L~,\label{eqn:KTM}\\
	\mathcal{J}_Z^\mu & = \frac{g}{2c_W}\bigg(1 + \frac{\kappa^2}{2}\bigg) \bar{\nu}_LU^\dagger U \gamma^\mu \nu_L - \frac{g\kappa^2}{2c_W}\bar{\nu}_L\gamma^\mu\nu_L~,\label{eqn:KTM2}
\end{align}
where we have defined the $3\times N$ leptonic mixing matrix, $U$, by
\begin{equation}
	\label{eqn:GFPM}
	U^{iI} = [V^{\ell\dagger}]^{ij}[V^n]^{jI}~.
\end{equation}
In Eq. (\ref{eqn:KTM2}) we have applied the identity $(V^\dagger)^{I\alpha}V^{\alpha J} = \delta^{IJ} - (V^\dagger)^{Ii}V^{iJ}$ and then inserted $V^\ell V^{\ell\dagger} = 1$ to write the non-diagonal $Z$ current in terms of $U^\dagger U$. It is worthwhile noting here that the unitarity of $V^{l}$ and $V$ implies
\begin{equation}
	[UU^\dagger]_{ij} = [V^{\ell\dagger}]_{ik}[V]_{kI}[V^{\dagger}]_{Im}[V^{\ell}]_{mj}  = \delta_{ij}~, \label{eqn:PMUR}
\end{equation}
whereas
\begin{equation}
	[U^\dagger U]_{IJ} = [V^{\dagger}]_{Ik} [V]_{kJ} \not= \delta_{IJ}~.
\end{equation}
Hence the mixing matrix is only unitary from the right. 

The Higgs couplings to the neutrinos are also non-diagonal in the mass basis. Explicitly, writing the massive Higgs as $h$, from Eqs. (\ref{eqn:LHFM}) and (\ref{eqn:MBN}) we have
\begin{equation}
	\mathcal{L}_h = h^\dagger \bar{N}_L^i \epsilon^{\tilde n} \tilde\lambda_{iI}N_R^I  = h^\dagger \bar{\nu}^I_L (V^{\dagger})^{Ii}\epsilon^{\tilde n}\tilde{\lambda}_{iJ} W^{JK} \nu^K_R~.
\end{equation}
From Eqs. (\ref{eqn:NMM}) and (\ref{eqn:BUD}) it follows that $(\theta \tilde{\lambda} W)^{iI} = (V d)^{iI}$, so then
\begin{equation}
	\label{eqn:HC}
	\mathcal{L}_h = h^\dagger\bar{\nu}_L \bigg(U^\dagger U\frac{m}{v}\bigg) \nu_R + \mbox{h.c}~,
\end{equation}
where $m$ is the diagonal neutrino mass matrix, $m \equiv \Lambda d$.

\subsection{Neutrino mass spectrum and eigenstates}
Let us now determine the neutrino mass spectrum, as well as the general structure of $V$.  We are particularly interested in determining $V$ since this matrix is a component of the leptonic mixing matrix $U$. The general structure of the unitary matrix $V$ can be determined simultaneously with the mass spectrum by diagonalizing $AA^\dagger$. Details of this diagonalization are presented in Appendix \ref{sec:NMBS}. One finds that the leading order general structure of $V$ is (\ref{eqn:AGSVN})
\begin{equation}
	\label{eqn:GSVN}
	V = \begin{pmatrix} X_3 & \theta W_K \\ \theta Y_3 & Z_K \end{pmatrix}~,
\end{equation}
where the matrices $X_3$, $Y_3$, $W_K$ and $Z_K$ have either $\mathcal{O}(1)$ entries or entries suppressed by powers of $\epsilon$. As a result of Eq. (\ref{eqn:GSVN}), the neutrino mass basis is a weak mixing between the SM fields $N_L^i$ and the heavy bound states $N_L^\alpha$, with mixing angle of order $\theta$.  A summary of the mass spectrum and corresponding eigenstates is presented in Table \ref{tab:NMSE}. Here and henceforth we denote the three light ($N-3$ heavier) neutrinos by $\nu_{L,R}^l$ ($\nu_{L,R}^H$).

\begin{table}[t]
\begin{tabular}{c|cc}
	\hline
	Eigenstate & Mass & Composition\\[3pt]
	\hline\hline
	$\nu_L^l$ & $\lesssim \theta \Lambda = v\epsilon^{\tilde n}$ & $\quad a_jN_L^j + \theta a_\alpha N_L^\alpha\quad$\\[3pt]
	$\nu_L^H$ & $\sim \Lambda$ or $\sim \epsilon^2\Lambda$ & $\quad a_j\theta N_L^j +  a_\alpha N_L^\alpha\quad$\\[3pt]
	\hline
\end{tabular}
\caption{Neutrino mass spectrum and composition of left-handed neutrino mass eigenstates.  The coefficients $a_j$ and $a_\alpha$ are $\mathcal{O}(1)$ numbers.}
\label{tab:NMSE}
\end{table}

\subsection{Charged and neutral currents}

The most significant consequence of this analysis is that since
$V^{\ell}$ has $\mathcal{O}(1)$ entries or smaller, then by
Eqs. (\ref{eqn:GFPM}) and (\ref{eqn:GSVN}) the mixing matrix has the
structure
\begin{equation}
	 \label{eqn:SPM}
	U  = \begin{pmatrix} V^{\ell\dagger}X_3 & \theta V^{\ell\dagger}W_K \end{pmatrix} \equiv \begin{pmatrix}  U_3 & \theta U_K \end{pmatrix}~,
\end{equation}
where $ U_{3,K}$ entries are either $\mathcal{O}(1)$ or suppressed by powers of $\epsilon$. In the mass basis the non-diagonal charged and neutral currents involving neutrinos in Eqs. (\ref{eqn:KTM}) and (\ref{eqn:KTM2}) are respectively
\begin{align}
	J^{\mu-}_{W} & = \frac{g}{\sqrt{2}}\Big(\bar{e}_L\gamma^\mu U_3\nu^l_L + \theta \big[\bar{e}_L \gamma^\mu U_K\nu^H_L\big]\Big)~,\notag\\
	J^\mu_Z & = \frac{g}{2c_W}\bigg[\bar{\nu}^l_L \gamma^\mu \Big(U_3^\dagger U^{\phantom{}}_3(1 + \kappa^2/2) - \kappa^2\Big)\nu^l_L +  \theta(1 + \kappa^2/2) \big[\bar{\nu}_L^l \gamma^\mu U_3^\dagger U^{\phantom{}}_K\nu^H_L  + \mbox{h.c.}\big] \notag\\
	& +  \bar{\nu}^H_L \gamma^\mu \Big(\theta^2U_K^\dagger U^{\phantom{}}_K(1 + \kappa^2/2) - \kappa^2\Big) \nu^H_L\bigg]~. \label{eqn:CNCN}
\end{align}
It is clear that the coupling of the heavy left-handed neutrinos $\nu^H_L$ to the $W$ and $Z$ is suppressed by either a $\theta$ or $\kappa^2$ factor. Similarly, writing $m = \mbox{diag}\{m_\nu,m_H\}$, the Higgs coupling (\ref{eqn:HC}) becomes
\begin{equation}
\label{eqn:HCMB}
	\mathcal{L}_h  = h^\dagger\bigg(\bar{\nu}_L^l U_3^\dagger U^{\phantom{}}_3\frac{m_\nu}{v}\nu_R^l + \epsilon^{\tilde n} \bar{\nu}_L^l U_3^\dagger U^{\phantom{}}_K\frac{m_H}{\Lambda}\nu_R^H + \epsilon^{\tilde n}\bar{\nu}_L^H U_K^\dagger U^{\phantom{}}_3 \frac{m_\nu}{\Lambda}\nu_R^l + \theta\epsilon^{\tilde n} \bar{\nu}_L^H U_K^\dagger U^{\phantom{}}_K\frac{m_H}{\Lambda}\nu_R^H\bigg)~,
\end{equation}
all the terms of which are suppressed by at least $\epsilon^{\tilde n}$, due to both the weak mixing with heavy mass eigenstates and the light neutrino masses.

\subsection{Scales and dark matter candidates}
Before continuing any further, let us examine possible scales for $\epsilon$. First,
note from Table \ref{tab:NMSE} that the light neutrino masses $m_\nu
\lesssim v\epsilon^{\tilde n}$. It must be that $\tilde{n} \ge 3$, and in particular let us assume one of the light neutrinos is comprised of three effective preons, so $\tilde {n} = 3$ and $m_\nu \sim v\epsilon^{3}$. The other two light neutrinos may be comprised of more effective preons, in which case their masses are further suppressed.
For $m_\nu \sim 0.1$ eV we then have
\begin{equation}
	\label{eqn:SDEN}
	\epsilon^{3} \sim \frac{m_\nu}{v} \sim 10^{-13}~\implies \epsilon \lesssim 10^{-4} ~.
\end{equation}
Further, from Eq. (\ref{eqn:DMALH}) it follows that the upper bound on the mixing between the light and heavy neutrinos
\begin{equation}
	\label{eqn:LHM}
	\theta \lesssim \frac{m_\nu}{\Lambda}~,
\end{equation}
and the secondary mass generation scale (if any) is, by Eq. (\ref{eqn:SDEN})
\begin{equation}
	M_{\textrm{2MG}} \sim \epsilon^{2}\Lambda \sim \bigg(\frac{m_\nu}{v}\bigg)^{2/3}\Lambda \sim 10^{-26/3}\Lambda~.
\end{equation}

A particularly interesting scenario is to consider $\Lambda \sim 1$ TeV, which is in line with the cosmological bounds mentioned in Ref. \cite{Okui:2004xn}. As a result, from Eq. (\ref{eqn:LHM}) the light-heavy mixing is $\theta \lesssim 10^{-13}$ and then the secondary mass generation scale
\begin{equation}
	M_{\textrm{2MG}} \sim \mbox{KeV}~.
\end{equation}
If there is secondary mass generation, then the mass and lifetime scales of such `intermediate' neutrinos, which we will
denote $\nu^m$, suggest they may be warm dark matter candidates \cite{Dodelson:1993je,Asaka:2005tn,Asaka:2006ek,Kusenko:2009up}. Evidence for the decay of such warm dark matter with a KeV mass scale has been potentially observed in the Willman 1 dwarf galaxy \cite{Loewenstein:2009cm}, although we note that the heavy light mixing angle for the intermediate neutrinos is much smaller than the claimed mixing in Ref. \cite{Loewenstein:2009cm}, which is $\sim 10^{-5}$. 

Explicitly, the $\nu^m$ are lighter than any massive content of the SM, so kinematically they may only decay into the light neutrinos.  The
leading order contribution to this decay is through the tree-level
neutral current $\nu^m \to 3\nu^l$ in Eq. (\ref{eqn:CNCN}).  From na\"\i ve
dimensional analysis the lifetime of these neutrinos is $\sim 10^{29}$
years.  Note also that these intermediate neutrinos could be produced
in a $W$ or $Z$ decay, but the contribution to the decay rate is
suppressed by at least $\theta^2 \sim 10^{-26}$ compared
to the contribution of the three light neutrinos, satisfying
experimental bounds on a fourth generation light neutrino coupled to
the massive SM gauge bosons \cite{PDG:2008pd}.  We also note from Eqs. (\ref{eqn:CNCN}) and (\ref{eqn:HCMB}) that since the coupling of the $\nu^H$ to the SM is generally suppressed by at least $\epsilon^{\tilde n} \sim 10^{-13}$ or $\kappa$, and they have masses $\sim$ TeV, then these heavy neutrinos may be candidate `feebly interacting massive particles' \cite{Hall:2009bx}.

\subsection{Non-unitarity and neutrino oscillations}
The large mass scale for the heavy neutrinos means that in a process such as
$\beta$-decay, production of heavy neutrino mass eigenstates $\nu^H_L$
is kinematically forbidden. As a consequence, the physical flavor
states produced in experiments will consist of combinations of only
the light $\nu^i_L$. That is, the neutrino flavor basis is defined by
\begin{equation}
	\label{eqn:NFB}
	n_{L}^{f} \equiv U_3^{f i}\nu^i_L~.
\end{equation}
Here the non-unitary $U_3$ replaces the $3\times 3$ unitary PMNS matrix found in the standard treatment of neutrino mixing (see e.g. \cite{PDG:2008pd}). We henceforth call $U_3$ the effective PMNS matrix. Note that intermediate neutrinos may also be produced, in which case the physical flavor states Eq. (\ref{eqn:NFB}) have further components suppressed by $\theta$: We can alternatively think of Eq. (\ref{eqn:NFB}) as defining the leading order terms of the flavor basis. 

The consequences of a non-unitary PMNS matrix have been previously examined in depth (see Refs \cite{Antusch:2006vwa,GiuntiEtAl:1992rw,Langacker:1988ln}, and references therein). For example, non-unitarity of the PMNS matrix means that the flavor states $|n^f_L\rangle$ are no longer orthogonal. In the context of neutrino oscillations, the probability $P_{f\to g}$ for a transition between states $|n^f_L\rangle$ and $|n^g_L\rangle$ is modified, and in particular there is a non-zero probability for a flavor transition at the source of a neutrino beam - a so-called zero distance effect. Other consequences are e.g. modified charged and neutral current cross-sections. 

In general, all modifications from the standard treatment due to non-unitarity arise from factors involving either $U_3U_3^\dagger$ or $U_3^\dagger U_3$. However, from Eqs. (\ref{eqn:PMUR}) and (\ref{eqn:SPM}) and the unitarity of $V$ (\ref{eqn:GSVN}), we have
\begin{align}
	U_3^{\phantom{}}U_3^\dagger & = 1 - \theta^2 U_K^{\phantom{}}U_K^\dagger~,\notag\\
	U_3^\dagger U_3^{\phantom{}} & = 1 - \theta^2 Y_3^{\dagger}Y_3^{\phantom{}}~,\label{eqn:U3U}
\end{align}
so that the unitarity of $U_3$ is only very weakly broken. That is, for our present low-energy theory the non-unitarity of $U_3$ only weakly modifies the standard formalism, including that of neutrino oscillations. For example, the probability associated with the zero distance effect
\begin{equation}
	P_{f \to g}(L=0)  = \frac{|(U^{\phantom{}}_3U_3^\dagger)_{fg}|^2}{(U^{\phantom{}}_3U_3^\dagger)_{ff}(U^{\phantom{}}_3U_3^\dagger)_{gg}}  \simeq \theta^4|(U_K^{\phantom{}}U_K^\dagger)_{fg}|^2 \sim 10^{-54}
\end{equation}
if $\Lambda \sim$ TeV, which is well below experimental threshold sensitivities. Similarly, the general expression for neutrino oscillation probabilities is the same as in the standard treatment, up to non-unitarity corrections at most $\mathcal{O}(\theta^2)$.

\section{Conclusion}
We have presented a mechanism through which light Dirac neutrinos may be naturally generated. An essential ingredient is the hidden flavor $\UF$ charge assigned to the Higgs scalar $\phi$, which produces an unbroken axial U(1) when combined with hypercharge. With particular hidden charge assignments to the fermionic fields, this axial symmetry forbids the production of Majorana masses during the spontaneous symmetry breaking induced by $\langle \phi \rangle$ alone. 

We have shown in this paper that a simple assignment of $\UF$ charges,
which is motivated in part by the need to reproduce the SM Yukawas
involving $\phi$, is non-anomalous and produces an unbroken axial
symmetry that is isomorphic to $B-L$, guaranteeing that only Dirac
fermions are produced.  In all this, the compositeness plays a key
role: it naturally produces a pattern of chiral symmetry breaking that
incorporates this extra symmetry into the theory and 
allows us to produce very light Dirac masses in
comparison to the rest of the SM fields. Further, compositeness also
naturally produces an arbitrary number of heavier neutrinos, which are
weakly coupled to the SM.

Our U(1) hidden flavor model predicts observable effects beyond the SM, though the effects in question
are not unique to our model. First, since it predicts Dirac neutrinos, no neutrino-less
double beta decay will be observed in our model.  Baryon and lepton number violating processes, the unitarity breaking of the effective PMNS matrix and the mixing between the light and heavy neutrino mass eigenstates are all strongly suppressed by compositeness, such that the effects associated with these features may be small enough to satisfy current experimental bounds. Phenomenology due to the extra gauge boson is similarly suppressed by its small coupling. Finally, the intermediate
neutrinos, whose masses may be produced by secondary mass generation, are potential warm dark matter candidates, while the weakly coupled heavier neutrinos could be so-called feebly interacting massive particles. 

\acknowledgments{The authors thank Nima Arkani-Hamed, Josh Berger, Bibhushan Shakya, Philip Tanedo and Yuhsin Tsai for helpful discussions. This work is supported by NSF grant number
PHY-0757868 and by a U.S.-Israeli BSF grant.}

\appendix
\section{Examples of Preonic Theories}
\label{sec:EPT}
In Sec. \ref{sec:UHFM} we presented a hidden flavor theory in which we assumed that there exists a $\Gc\otimes\GF$ effective preonic theory that produces three chiral bound states all with the same hidden flavor charge, but possibly composed of different numbers of effective preons.  This assumption is non-trivial, so in this appendix we search for examples of effective preonic theories which possess this feature. We emphasize that the following theories are just examples of theories with the required group theoretic structure to produce three chiral bound states subject to the anomaly matching conditions: we do not intend to imply any of them are actually the effective preonic theory. For brevity, in the remainder of this appendix we will contract `effective preon' to just `preon'.

The $\SU(n+4)\otimes \SU(n)\otimes U(1)$ preonic theories considered
in Refs
\cite{ArkaniHamedGrossman:1999,DimopoulosRabySusskind:1980,Okui:2004xn}
produce an effective low-energy theory after only one stage of
confinement. A good place to start is therefore with a preonic theory
that has the same confining group representations as these
theories. Hence consider a preonic theory with symmetries $\Gc =
\SU(n)$ and $\GF = G\otimes \UF $, and preonic content as shown in
Table \ref{tab:PTFC}. Here $n\ge 5$ and the group $G$ is semi-simple
but arbitrary: the representations of $G$ furnished by the preons are
specified only by their dimensions $d_1$ and $d_2$. We assume that $G$
is spontaneously broken by confinement, so that the low energy theory
has only a $\UF$ flavor symmetry. As a result, only the SU(n)$^2\UF$
instanton and $\UF^3$ anomalies need to be matched. Note that in
contrast to the main text, for convenience we have switched to a
right-handed chirality for the preonic representations. The anomalies
in this section will be calculated with respect to the right-handed
fermionic representations, and therefore differ by a sign compared to
those in Sec. \ref{sec:UHFM}.
\begin{table}[t]
\begin{tabular}{c|ccc}
	\hline Field & $\quad \SU(n)\quad$ & $\quad G\quad$ & $\quad
	\UF \quad$\\[3pt] \hline\hline $\psi$ & $\fund$ & $d_1$ &
	$\alpha$\\[3pt] $\chi$ & $\abfund$ & $d_2$ & $\beta$\\[6pt]
	\hline
\end{tabular}
\caption{Right-handed fermionic content for a candidate $\Gc\otimes\GF$ preonic theory. }
\label{tab:PTFC}
\end{table}

\subsection{Statistical, group theoretic and chiral constraints}
The effective theory after confinement consists of $\Gc$ singlets. We denote a general $\Gc$ singlet by $\psi^p\chi^q$, which has $\UF$ charge
\begin{equation}
	F(\psi^p\chi^q) = p\alpha + q\beta~.
\end{equation}
Note that in this notation, a negative power $\psi^{-|p|} \equiv (\psi^\dagger)^{|p|}$. In order for $\psi^p\chi^q$ to be both an $\SU(n)$ singlet and a fermion, the integers $p$ and $q$ must satisfy the respective constraints
\begin{align}
	p + (n-2) q \!\!\!\mod n  & = 0~, \label{eqn:PQSC}\\
	p + q \!\!\!\mod 2 & = 1~. \label{eqn:PQFC}
\end{align}
We also require this spin-1/2 bound state to be right-handed, which means that the
condensate $\psi^p\chi^q$ must contain an odd number of right-handed
preons and an even number of left-handed ones. Since by construction
$\psi$ and $\chi$ were right-handed -- so that $\psi^\dagger$ and
$\chi^\dagger$ are left-handed -- there are only four possibilities for
the configuration of the sign and parity of $p$ and $q$ which satisfy
this constraint. These are as shown in Table \ref{tab:PQSP} and are
equivalent to the algebraic constraint
\begin{equation}
	\label{eqn:PQLC}
	\big[1 + \mbox{sgn}(p)](p \!\!\!\mod 2) + \big[1 + \mbox{sgn}(q)] ( q \!\!\! \mod 2) = 2~.
\end{equation}
Notably, this constraint subsumes Eq. (\ref{eqn:PQFC}).
\begin{table}[t]
\begin{tabular}{cc|cc}
		\hline
		$\quad$ Sign $p\quad$ & $\quad$ Sign $q\quad$ & $\quad$ Parity $p\quad$ &$\quad$  Parity $q\quad$\\
		\hline\hline
		$+$ & $+$ & odd& even\\
		$+$ & $+$ & even & odd\\
		$+$ & $-$ & odd & even \\
		$-$ & $+$ & even & odd \\
		\hline
\end{tabular}
\caption{Four possible configurations of signs and parity for $p$ and $q$ that produce right-handed spin-1/2 bound states.}
\label{tab:PQSP}
\end{table}

\subsection{$\UF$ anomaly matching}

We now apply the 't~Hooft anomaly matching formalism. To begin, we note that the $\SU(n)^3$ anomaly for this theory must cancel.  This implies that
\begin{equation}
	\label{eqn:SUN3}
	d_1= (n-4)d_2~.
\end{equation}
Further, the preonic theory must have no $\SU(n)^2\UF$ instanton anomaly, which together with Eq. (\ref{eqn:SUN3}) results in
\begin{equation}
	\label{eqn:SUNI}
	 \alpha = \frac{2-n}{n-4}\beta \equiv g(n)\beta~.
\end{equation}
Combining Eqs. (\ref{eqn:SUN3}) and (\ref{eqn:SUNI}) we have $\UF^3$ anomaly
\begin{align}
	\mathcal{A}[\UF^3]
	& = \alpha^3nd_1 + \beta^3 n(n-1)d_2/2\notag\\
	& = \frac{nd_2\beta^3}{(n-4)^2}\bigg[(2-n)^3 + \frac{(n-1)(n-4)^2}{2}\bigg]\notag\\
	& \equiv f(n,d_2)\beta^3~.\label{eqn:UF3}
\end{align}
	
Now, let us suppose that there are precisely three right-handed chiral bound states formed from this preonic theory, which all have hidden flavor charge $\gamma$. Then the $\UF^3$ anomaly of the confined phase is simply $3\gamma^3$, so by 't~Hooft anomaly matching and Eq. (\ref{eqn:UF3}) we must have
\begin{equation}
	\label{eqn:UF3AM}
	f(n,d_2)\beta^3 = 3\gamma^3~.
\end{equation}
If we assume that there is at least one combination of preons which forms a spin-1/2 bound state of charge $\gamma$, then there must exist integers $p$ and $q$ such that  $\gamma = p\alpha + q\beta$. Observe that by Eq. (\ref{eqn:SUNI})
\begin{equation}
	\label{eqn:GBQ}
	\gamma/\beta = p \alpha/\beta + q = p g(n) + q~,
\end{equation}
so it follows that a necessary condition for the anomaly matching constraint (\ref{eqn:UF3AM}) to be satisfied is
\begin{equation}
	\label{eqn:PQAMC}
	f(n,d_2) = 3[p g(n) + q]^3~.
\end{equation}
The reason Eq. (\ref{eqn:PQAMC}) is not a sufficient condition is because we require there to be three chiral bound states, but it is conceivable that if $p$ and $q$ are small enough, then there may not be enough spin-1/2 bound states formed from this combination alone: this depends on the multiplicities generated by both the broken group $G$ as well as tensor products of the Lorentz indices. Interestingly, note that by Eq. (\ref{eqn:GBQ}) $\gamma/\beta$ is always rational and so Eq. (\ref{eqn:UF3AM}) also implies the severe constraint
\begin{equation}
	\label{eqn:NDAMC}
	\bigg(\frac{f(n,d_2)}{3}\bigg)^{1/3} \in \mathbb{Q}~.
\end{equation}
This provides a necessary constraint on the combinations of $n$ and $d_2$ such that there may be three chiral bound states of the same charge for the class of preonic theories defined by Table \ref{tab:PTFC}. Note that Eq. (\ref{eqn:PQAMC}) implies Eq. (\ref{eqn:NDAMC}), but the converse does not hold. 

\subsection{Gravitational anomaly matching}

In order for a preonic theory of this class to generate three chiral bound states with the same charge, we must find integers $n \ge 5$, $d_2\ge 1$, $p$ and $q$ which satisfy Eqs. (\ref{eqn:PQSC}), (\ref{eqn:PQLC}) and (\ref{eqn:PQAMC}).  One further constraint is produced by the fact that the gravitational anomaly must also match. 

The gravitational anomaly for the spin-1/2 bound states is $3\gamma$. For the preonic theory we have
\begin{equation}
	\mathcal{A}_{\textrm{grav}}[\UF]  = nd_1\alpha + n(n-1) d_2 \beta/2 = \frac{n d_2}{2} (3-n)\beta~.
\end{equation}
From Eq. (\ref{eqn:UF3AM}), simultaneous matching of the gravitational and $\UF^3$ anomalies then requires that
\begin{equation}
	f(n,d_2) = 3\bigg[\frac{n d_2}{6} (3-n)\bigg]^3~.
\end{equation}
The only integer solutions for this equation with $n \ge 5$ are $n=5$, $d_2 = 3$ or $n=6$, $d_2 = 1$. A computer search reveals that for these values, there exists a large number of $p$ and $q$ configurations which satisfy Eqs. (\ref{eqn:PQSC}), (\ref{eqn:PQLC}) and (\ref{eqn:PQAMC}). These theories are presented in Tables \ref{tab:CPTC1} and \ref{tab:CPTC2}, along with some possible bound state configurations. Examples of preonic theories with such field content are respectively three copies of the $\SU(5)\otimes U(1)$ theory or the $\SU(6)\otimes \SU(2)\otimes U(1)$ theory that are presented in Ref. \cite{DimopoulosRabySusskind:1980}. With regard to the latter example, please note that we need not generally set $G = \SU(2)$ in Table \ref{tab:CPTC2}: We could also simply have two copies of the $\psi$ field and $G$ just a multiplicity.

\begin{table}[t]
	\begin{tabular}{c|ccc}
		\hline
		Preon & $\quad \SU(5)\quad$ & $\quad G\quad$ & $\quad \UF\quad$  \\
		\hline\hline
		$\psi$ & $\fund$ & $d_1 = 3$ & $\alpha = 3\gamma/5$  \\[3pt]
		$\chi$ & $\abfund$ & $d_2 = 3$ & $\beta = -\gamma/5$  \\[6pt]
		\hline
		Singlet &$\quad \SU(5)\quad$ &  &$\quad \UF \quad$ \\
		\hline\hline
		$\psi^2\chi$ & $\bm{1}$ &  & $\gamma$ \\
		$\psi(\chi^\dagger)^2$ & $\bm{1}$ &  & $\gamma$ \\
		$\psi^3\chi^4$ & $\bm{1}$ &  & $\gamma$ \\
		$\vdots$&&&\\
		\hline
	\end{tabular}
	\caption{Preonic field content and possible chiral bound states for the candidate \SU(5) preonic theory.}
	\label{tab:CPTC1}
\end{table}
\begin{table}[t]
	\begin{tabular}{c|ccc}
		\hline
		Preon & $\quad \SU(6)\quad$ & $\quad G\quad$ & $\quad \UF \quad$  \\
		\hline\hline
		$\psi$ & $\fund$ & $d_1 = 2$ & $\alpha = 2\gamma/3$  \\[3pt]
		$\chi$ & $\abfund$ & $d_2 = 1$ & $\beta = -\gamma/3$  \\[6pt]
		\hline
		Singlet &$\quad \SU(6)\quad$ & &$\quad \UF \quad$ \\
		\hline\hline
		$\psi^2\chi$ & $\bm{1}$ &  & $\gamma$ \\
		$\psi^4\chi^5$ & $\bm{1}$ &  & $\gamma$ \\
		$\psi^6\chi^9$ & $\bm{1}$ &  & $\gamma$\\
		$\vdots$&&&\\
		\hline
	\end{tabular}
\caption{Preonic field content and possible chiral bound states for the candidate \SU(6) preonic theory.}
\label{tab:CPTC2}
\end{table}
It is clear that there is sufficient $p,q$ combinations to produce
three right-handed chiral bound states with $\UF$ charge $\gamma$ (not
including multiplicities from broken $G$ reresentations or tensor
products of Lorentz representations).  Note also that neither of these
theories can exhibit secondary mass generation. The reason is that all
the spin-1/2 bound states of charge $\gamma$ must contain a common preon with any
scalar condensate: for secondary mass generation we require more
sophisticated preonic content.

We have thus found two $\Gc\otimes\GF$ preonic theories which can produce three right-handed chiral bound states of the same $\UF$ charge. Of course, there exists many more possibilties which we have not considered here. We have also not specified here the pattern of chiral symmetry breaking, or exactly which of the above condensates correspond to the chiral bound states. These will depend on the ultraviolet completion of the theory, the broken hidden flavor symmetry $G$, the dynamics of confinement, and other physical assumptions, which is beyond the scope of the present paper.

\section{Gauge boson structure and couplings}
\label{sec:GBSC}
In this appendix we present the gauge boson mass basis and further detail of the gauge boson couplings. 

With an extra $\UF$ gauge symmetry we have covariant derivative
\begin{equation}
 	iD_\mu = i\partial_\mu  - gT^aW^a_\mu  - g^\prime Y B_\mu  - g_{\textrm{F}} F C_\mu~,
\end{equation}
where as usual $T^a$ ($W^a_\mu$) are the $\SU(2)_L$ generators (gauge bosons), and $C_\mu$ is the $\UF$ gauge boson. Define
\begin{align}
	\cos\theta_F & \equiv \frac{2\gamma g_{\textrm{F}}}{\sqrt{g^2 + g^{\prime 2} + (2\gamma g_{\textrm{F}})^2}}~,\notag\\
	\cos\theta_W & \equiv \frac{g}{\sqrt{g^2 + g^{\prime 2}}}~,
\end{align}
and let us write $\cos\theta_{F,W} = c_{F,W}$, $\sin \theta_{F,W} = s_{F,W}$. It is straightforward to find the orthonormal mass basis for the gauge bosons, which is presented in Table \ref{tab:GBMB}. Note that the massless gauge bosons $A_\mu$ and $A^\prime_\mu$ are orthonormal with respect to the $\{W^3,B,C\}$ basis. 

\begin{table}[t]
\begin{tabular}{c|cc}
	\hline
	Boson & Structure & Mass$^2$ \\[3pt]
	\hline\hline
	$W_\mu^{\pm}$& $\quad(W_\mu^1 \mp iW_\mu^2)/\sqrt{2}\quad$ & $v^2g^2/2$\\[3pt]
	$Z_\mu$ &  $\quad c_Ws_F W_\mu^3 - s_Ws_F B_\mu - c_F C_\mu \quad$ & $v^2g^2/2 s^{2}_Fc^{2}_W$\\[3pt]
	$A_\mu$ & $\quad s_WW^3_\mu + c_W B_\mu\quad$ & 0\\[3pt]
	$A_\mu^\prime$ & $\quad c_Wc_F W_\mu^3 - s_Wc_F B_\mu + s_F C_\mu \quad$ & 0 \\[3pt]
	\hline
\end{tabular}
\caption{Gauge boson mass basis for the U(1) hidden flavor model.}
\label{tab:GBMB}	
\end{table}

We are generally free to choose orthonormal $A_\mu$ and $A^\prime_\mu$ up to a unitary transformation. However, the choice of the basis for the massless gauge bosons in Table \ref{tab:GBMB} proves to be particularly convenient, because in this mass basis the covariant derivative becomes
\begin{align}
	iD_\mu
	& = i\partial_\mu -gT^{\pm}W^{\pm}_\mu  - eQA_\mu  - \frac{g}{c_W}\bigg[s_F\big(T^3 - Qs_W^2\big) - \bigg( Y - \frac{B-L}{2}\bigg)\frac{c^2_F}{s_F}\bigg]Z_\mu\notag\\
	& - \frac{gc_F}{c_W}\bigg[Qc_W^2 - \frac{B-L}{2}\bigg]A^\prime_\mu~,
\end{align}	
where $e = gs_W$ and we have used the relations $F/2\gamma = Y - a/\gamma$ and $a/\gamma = (B-L)/2$ if $L=1$ for the electron.  It is clear that $A_\mu$ is the SM photon. Note that the generator of $U(1)^\prime$, whose gauge boson is the $A_\mu^\prime$, is a linear combination of $Q$ and $B-L$.

Consider now the limit $g_{\textrm{F}} \ll g, g^\prime$. We can redefine this limit as
\begin{equation}
	\label{eqn:AKD}
	\kappa \equiv c_F \ll 1~,
\end{equation}
so that $s_F = 1 - \kappa^2/2 + \mathcal{O}(\kappa^4)$. To order $\mathcal{O}(\kappa^2)$, we then have mass basis gauge bosons
\begin{align}
	Z_\mu & \simeq (1 - \kappa^2/2) Z^{\textrm{SM}}_\mu + \kappa C_\mu~,\notag\\
	A_\mu^\prime & \simeq \kappa Z^{\textrm{SM}}_\mu + (1 - \kappa^2/2) C_\mu~,
\end{align}
where $Z^{\textrm{SM}}$ is the SM Z boson, and $Z_\mu$ has mass $m_Z^2 \simeq (1 + \kappa^2/2)v^2g^2/2c_W^{2}$. Further, the covariant derivative in this limit is simply
\begin{align}
	iD_\mu
	& \simeq i\partial_\mu -gT^{\pm}W^{\pm}_\mu  - eQA_\mu - \frac{g}{c_W}\bigg[ \big(T^3 - Qs_W^2\big) - \frac{\kappa^2}{2}\big(Qc_W^2 +Y - B-L\big)\bigg]Z_\mu\notag\\
	& - \frac{g\kappa}{c_W}\bigg[Qc_W^2 - \frac{B-L}{2}\bigg]A^\prime_\mu~.\label{eqn:ACDWGL}
\end{align}
Hence at leading order in $\kappa$ we have the usual SM gauge bosons, mass spectra and couplings.

\section{Neutrino mass basis and spectrum}
\label{sec:NMBS}
Consider 
\begin{equation}
	AA^\dagger = \begin{pmatrix} \theta^2(\tlt\tlt^\dagger + \tlK\tlK^\dagger) & \theta\tlK d_{K} \\ \theta d_{K} \tlK^\dagger & d^2_{K}\end{pmatrix}~.
\end{equation}
The upper left $3\times 3$ block is Hermitian, and can be diagonalized by unitary $V_3$, so that we then have
\begin{equation}
	AA^\dagger = \begin{pmatrix} V_3 &\\&1\end{pmatrix} \!\!\begin{pmatrix} \theta^2  d_3^2 & \theta V_3^\dagger\tlK d_{K} \\ \theta d_{K} \tlK^\dagger V_3 & d^2_{K}\end{pmatrix}\!\!\begin{pmatrix} V_3^\dagger &\\&1\end{pmatrix}~. \label{eqn:AAD}
\end{equation}
Since $\tilde\lambda$ contains at least one $\mathcal{O}(1)$ column, we expect at least one entry of the diagonal $d_3$ to be $\mathcal{O}(1)$ too, while the others may be suppressed. The exact nature of $d_3$, including any hierarchies therein, strongly depends on the structure and symmetries (if any) of $\tilde\lambda$. Determining $\tilde\lambda$ is beyond the scope of this paper.

To leading order in $\theta$, one may show that the characteristic equation for $AA^\dagger$ is
\begin{equation}
	0  = \prod_{i=1}^3(\theta^2d^2_i - x)\prod_{\alpha = 4}^N(d^2_\alpha - x) - \theta^2\sum_{j,\beta}\Big[\prod_{i\not=j}\prod_{\alpha\not=\beta}(d^2_\alpha - x)(\theta^2d^2_i-x)B_{j\beta}B_{j\beta}^*\Big]~, \label{eqn:CEAA}
\end{equation} 
where $B = V_3^\dagger\tlK d_K$, $d_i = [d_3]_{ii}$ and $d_\alpha = [d_K]_{\alpha\alpha}$. It follows immediately from this and Eq. (\ref{eqn:LLY}) that the mass spectrum for the neutrinos is
\begin{align}
	m_i & =  v\epsilon^{\tilde n} d_i \big[1+ \mathcal{O}(\theta^2)\big]~,\notag\\
	m_\alpha & = \Lambda d_\alpha \big[1+ \mathcal{O}(\theta^2)\big]~.
\end{align}
Hence we have three light neutrinos, with masses suppressed by at least $\epsilon^{\tilde n}$ compared to the charged leptons, and $N-3$ neutrinos with masses $\sim \Lambda$ (or $\sim \Lambda \epsilon^{2}$). It follows from Eqs (\ref{eqn:AAD}) and (\ref{eqn:CEAA}) that the leading order general structure of $V$ must be
\begin{equation}
	\label{eqn:AGSVN}
	V = \begin{pmatrix} X_3 & \theta W_K \\ \theta Y_3 & Z_K \end{pmatrix}~.
\end{equation}


%

\end{document}